\documentclass[12pt]{iopart}

\usepackage{graphicx}
\usepackage{amsfonts, amssymb}
\usepackage{bbm}
\usepackage[perpage]{footmisc}

\usepackage{hyperref}

\hypersetup{
   colorlinks=true,         
   linkcolor=blue,          
   citecolor=red,        
   urlcolor=Violet            
}

\newcommand{\id}{\mathbbm{1}}
\newcommand{\ot}{\otimes}

\newcommand{\eq}[1] {(\ref{#1})}

\begin{document}

\title{Relative Locality in $\kappa$-Poincar\'e}

\author{Giulia Gubitosi$^{1,2}$ and Flavio Mercati$^{3,4}$}
\address{
$^1$Berkeley Lab \& University of California, Berkeley, 
CA 94720, USA.\\
$^2$Dipartimento di Fisica and sez. Roma1 INFN,
Universit\`a di Roma La Sapienza, P.le A. Moro 2, 00185 Rome, Italy.\\
$^3$Departamento de F\'isica T\'eorica, Universidad de Zaragoza, Zaragoza 50009, Spain.\\
$^4$Perimeter Institute for Theoretical Physics, 31 Caroline Street North  Waterloo, ON N2L 2Y5, Canada.}

\begin{abstract}
We show that the $\kappa$-Poincar\'e  Hopf algebra can be interpreted in the framework of curved momentum space leading to  relative locality.\\
We study the geometric properties of the momentum space  described 
by $\kappa$-Poincar\'e, and derive the consequences for particles propagation
and energy-momentum conservation laws in interaction vertices,
obtaining for the first time a coherent and fully workable model 
of the deformed relativistic kinematics implied by $\kappa$-Poincar\'e.\\
We describe the action of boost transformations on multi-particle systems, showing that in order to keep covariant the composed momenta it is necessary to introduce a dependence of the rapidity parameter on the particles momenta themselves.
Finally, we show that this particular form of the boost transformations keeps the validity of the relativity principle, demonstrating the invariance of the equations of motion under boost transformations.
\end{abstract}

\maketitle

\section{Introduction}

The problem of quantum gravity is one of the most elusive in modern physics.
It has been  repeatedly suggested that radical new ideas might be needed to
tackle it, in particular, we might be approaching the limit of applicability of 
Riemannian geometry in the description of spacetime.
A ``bottom-up'' approach to this problem might begin by attempting to 
describe some particular regimes in which the quantum properties
of the geometry of spacetime are under control.
A very interesting regime is the one in which all the gravitational degrees
of freedom are integrated away, leading to an effective field theory
for the matter fields. Of course this can't be done explicitly for the
full quantum theory of gravity, but in \cite{FreidelLivine} it has been done in 2+1 dimensions,
where gravity can be quantized as a topological field theory
and can be coupled to point particles, represented by topological
defects. Interestingly, the effective theory describing the dynamics of the
particles after integrating away the gravitational degrees of freedom
doesn't look like the dynamics of particles moving on a background 
spacetime manifold: such a situation is recovered only in the
low-energy limit. The matter degrees of freedom are representations
of an algebraic structure known as $\kappa$-Poincar\'e group \cite{LukierskiInventsKpoincare2,LukierskiInventsKpoincare1,MajidRueggBicross}
(in the 2+1dimensional version). The $\kappa$ refers to an energy constant, setting the scale at which the effective description of quantum-gravity effects provided by the model should begin to break down. Its role is analogous to that
of the Fermi constant $G_F$ in particle physics, which represents the only coupling of the effective
theory of weak interaction formulated by Fermi, and is today understood as coming from a 
deeper theory, whose constants ($g$ and $m_W$) combine to form $G_F$.
$\kappa$-Poincar\'e is a ``quantum deformation'' of the Poincar\'e group,
making it into a so-called ``quantum group'' or Hopf algebra \cite{Woronowicz},
a structure mathematicians proposed a few decades ago as the geometrical
tool to describe the symmetries of a noncommutative space \cite{majidbook}.
The analysis of  \cite{FreidelLivine} can't be repeated, at this stage, in the more
physical 3+1-dimensional case, but the $\kappa$-Poincar\'e group can be
easily generalized to arbitrary dimensions, and therefore it becomes an
interesting tool that is expected to capture some essential features of 
the low-energy limit of quantum gravity.

A recent series of works \cite{AFKS,AFKSfqxi}  proposed a new framework, called ``relative
locality'', in which to understand the physics of  a quantum gravity regime characterized by negligible $\hbar$ and $G$.
In this regime both quantum and gravitational effects are small, but the limits $\hbar \to 0$ and $G\to 0$
are taken so that their ratio is kept fixed, and we still have an energy scale $M_{p}\sim\sqrt{\hbar/G}$
governing modifications of standard physics. In relative locality the fundamental notion is that of momentum
space, which is a (pseudo) Riemannian manifold,
which might be curved and have other nontrivial geometrical properties, such as torison and non-metricity.
Space and time, on the other hand, loose their geometrical status. In particular, the notion
of locality becomes observer-dependent: the  fact that two events take place at the same spacetime 
point is not an absolute concept, and can only be established by observers close to the events themselves (relativity of locality). The relative locality proposal resulted from a deepening in the understanding of the fate of the locality principle in quantum gravity-motivated generalizations of Special Relativity \cite{FlavioPRL,SabineLocality,LeeOnLocality1,LeeOnLocality2}.

The relation between the ``relative locality'' regime and the one considered in \cite{FreidelLivine}  is apparent,
and therefore it's natural to explore the relationship between the relative locality
framework and $\kappa$-Poincar\'e.
Interestingly, several authors in the last decade have suggested the interpretation of the group manifold underlying
$\kappa$-Poincar\'e as a curved momentum space (e.g. \cite{dSKow02,dSKow03,dSKow04}), but the 
physical meaning of that is still unclear.

In this paper we apply al the machinery developed in \cite{AFKS} to the case of $\kappa$-Poincar\'e, identifying
the geometrical properties of the momentum space it describes, such as the metric, curvature, torsion and non-metricity, which reflect into different kinematical and dynamical properties of the motion and
interaction of particles. 
This construction allows us to deduce the physical implications of $\kappa$-Poincar\'e in a simple model whose physical interpretation is clear, which is what has been missing the most since the discovery of this Hopf algebra.

%
%
 
 
In the next Section we briefly review the physical implications of the geometrical properties of momentum space emerging in the Relative Locality framework.

In Section 3 we show how the translation sector of the $\kappa$-Poincar\'e Hopf algebra can be used to represent the coordinates over a momentum space, establishing a
general correspondence between commutative Hopf algebras and the geometric
structures introduced in Ref. \cite{AFKS}, in a way that can be applied also to other Hopf algebras. 

In the following Section 4 
we review the construction of $\kappa$-Poincar\'e as
a momentum space with de Sitter metric and with torsion and nonmetricity. We can then follow the prescriptions given in Section 2 to make the connection between the geometrical properties of this momentum space and the physics that it describes.
A byproduct of our analysis is the identification of a 
dispersion relation
that is natural from the perspective of the geometry of momentum space
(it is the geodesic distance from the origin), and such that the mass corresponds to the particle's rest energy.

%
%
Within this interpretation of  $\kappa$-Poincar\'e it is possible to show  (and we do this in Section 5)  that Lorentz transformations act nonlinearly on momenta, and, even more curiously,
that they have to act differently on different momenta, when they belong to an interaction
vertex, in order to keep covariant the total momentum of the vertex. In particular, we find that
that different particles participating to an interaction vertex transform under Lorentz transformation with different rapidities, which depend on the momenta of the other particles involved.

%
%
 
In section 6 we show  that  in this physical framework  the relativity principle still holds, in the sense that the equations of motion are invariant under boost transformations.
This result is particularly relevant because in the past
the issue of whether $\kappa$-Poincar\'e implies a breakdown of the
relativity principle was subject to debate \cite{Meljanac}. Of course, the interest of $\kappa$-Poincar\'e as an algebra of
physical symmetries would be seriously reduced, if it turned out that it
implied the breakdown of those symmetries. Our result provides the 
first explicit example of how  the 
equivalence between inertial observers is realized in the context of Relative Locality,
and it turns out to be realized in a particularly nontrivial way.

%


In Section 7 we identify a structure, related to the tangent space at the origin
of momentum space, which reproduces the commutation relations
of $\kappa$-Minkowski, a noncommutative spacetime whose symmetries are
thought to be described by $\kappa$-Poincar\'e \cite{MajidRueggBicross}. This result 
suggests that such a noncommutative space could emerge upon quantization
of certain (space-time) degrees of freedom of our model.


\section{Preliminaries on the Relative Locality principle}

The relativity of locality is achieved in \cite{AFKS} by assuming the phase space as
the fundamental arena where physics takes place, considered
as the cotangent bundle to momentum space.
Momentum space is assumed to be a pseudo-Riemannian
manifold $\Sigma$ which possess a distinguished
point $\underline 0$ (the \emph{origin}), a \emph{metric} $g$ and a 
\emph{connection} $\Gamma$, which doesn't necessarily need to be metric.

Physical observables are related to intrinsic geometric concepts.
The mass of a particle is measured by the geodesic distance of
the particle's representing point in momentum space from the origin,
\begin{equation}
d^2(p,\underline{0}) = m^2 ~.\label{eq:DispRelAFKS}
\end{equation}
This  equation gives the dispersion relation. In this sense  the metric of momentum space is related to the kinematical properties of a single particle.\footnote{ Note  that the dispersion relation 
depends on the particular choice of coordinate system over the momentum
space. }

Dynamics, or the interaction between particles, is related to the connection, since the connection defines the composition law of momenta,
$
\oplus : \Sigma \times \Sigma \to \Sigma$ \footnote{This law is assumed to admit a left and right inverse $\ominus : \Sigma \to \Sigma$, such that
$
\ominus p \, \oplus p = p \oplus \,( \ominus p) = \underline{0} .
$},
%
through
\begin{equation}
\left. \frac{\partial}{\partial p_\mu} \frac{\partial}{\partial q_\nu} (p \oplus_k q)_\rho \right|_{p=q=k} = - \Gamma^{\mu\nu}_\rho (k)~, \label{ConnectionAtAPoint}
\end{equation}
where $\oplus_k$ is the composition law ``translated'' at the point $k$
\begin{equation}
p \oplus_k  q = \oplus k \left( (\ominus k \, \oplus p) \oplus (\ominus k \, \oplus q)  \right) ~. 
\label{TranslatedComposition}
\end{equation}

The antisymmetric part of the connection is the torsion, which measures the noncommutativity of the composition law
\begin{equation}
\left. \frac{\partial}{\partial p_{\mu}} \frac{\partial}{\partial q_{\nu}} (p \oplus_k q - q \oplus_k p)_\rho \right|_{p=q=k} = - T^{\mu\nu}_\rho (k) ~, \label{eq:torsionAFKS}
\end{equation}
while the curvature measures its nonassociativity
\begin{equation}
\left. \frac{\partial}{\partial p_{[\mu}} \frac{\partial}{\partial q_{\nu]}}\frac{\partial}{\partial r_{\rho}} \left((p \oplus_k q) \oplus_k r - p \oplus_k (q \oplus_k r) \right)_\sigma \right|_{p=q=r=k} = R^{\mu\nu\rho}_\sigma (k) ~.\label{eq:CurvatureDef}
\end{equation}

The nonmetricity, defined from the metric and the connection as
\begin{equation}
N^{\mu\nu\rho} = \nabla^\rho g^{\mu\nu} (k) ~, \label{defNonmetricity}
\end{equation}
has been shown \cite{AFKS,LeeLaurentGRB} to be responsible for the leading order time-delay effect in the arrival of photons from distant sources, which is an effect
that is currently under experimental verification \cite{GiovanniLeeProspectsGRB}.

The dynamics of interacting particles is obtained from a variational principle. In the case of a single
vertex (interaction among $n$ particles with momenta $p^{j}, j=1...n$)  we need to minimize the following action:
\begin{eqnarray}
S &=&\sum_j \Bigg[\pm \int_{\sigma_0}^{\pm \infty} d \sigma\Bigg( \left\{ - x^\mu_j \dot{p^j}_\mu + N_j \left( d^2(p^j,\underline{0}) - m^2 \right)  \right\}\nonumber\\ 
&&\qquad\qquad+ z^\mu \, \mathcal{K}_\mu(p^1(\sigma_0) ,\dots, p^n(\sigma_0))\Bigg)\Bigg] ~.
\end{eqnarray}
The $\pm$ sign is chosen according to whether the $j$-th particle is outgoing or incoming.
$N_j$ and $z^\mu$ are Lagrange multipliers, but $z^\mu$ gives also the coordinates of the interaction point. $\mathcal{K}_\mu(p^1(\sigma_0) ,\dots, p^n(\sigma_0)) $ may be any combination of all of the momenta in the vertex and gives the momentum
conservation law, performed with the rules $\oplus$ and $\ominus$.~
$x^\mu_j$ are the spacetime coordinates of the $j$-th particle, and the dot represents
the derivative with respect to $\sigma$, an unphysical variable that parametrize the trajectory
of the system in phase space. $\sigma_0$ is an arbitrary value of $\sigma$ at which the
interaction is assumed to take place.

The constraints given by the variation with respect to the Lagrange multipliers $N_n$ and $z^\mu$ are
\begin{equation}
d^2(p^j,\underline{0}) = m^2 ~, \label{constraint1}
\end{equation}
which is the dispersion relation, and
\begin{equation}
\mathcal{K}_\mu(p^1(\sigma_0) ,\dots, p^n(\sigma_0)) = 0 ~, \label{constraint2}
\end{equation}
which gives the conservation of energy and momentum in the interaction vertex.

The (bulk) equations of motion resulting from the minimization of the action are
\begin{equation}
 \qquad \dot{p^j}_\mu = 0~, ~~ \dot {x^\mu}_j = - N_j \frac{\partial}{\partial p^j_\mu} d^2(p^j,\underline{0}) ~. \label{BulkEquations}
\end{equation}
The first equation expresses the conservation of particle momenta during
free propagation. The second one implies that the spacetime worldlines are straight lines, 
and their speed is $v_k = \frac{\partial}{\partial p^j_k} d^2(p^j,\underline{0}) / \frac{\partial}{\partial p^j_0} d^2(p^j,\underline{0})$. In the case of special relativity 
the angular coefficient is the relativistic speed
$v_k = p_k / \sqrt{p^2 + m^2}$. Note that the Lagrange multiplier $N_j$ simply amounts
to a normalization constant for the tangent vector to the trajectory, and has no
physical meaning.

The boundary terms give the initial conditions
\begin{equation}
x_j^\mu(\sigma_0)  = z^\mu \frac{\partial}{\partial p^j_\mu} \mathcal{K}_\mu(p^1,\dots p^n)~.
\label{BoundaryEquations}
\end{equation}
In the case of special relativity $\mathcal{K}_\mu(p^1,\dots p^n) = \sum_j p_\mu^j$ and all the worldlines simply end up at the interaction point $z^\mu$. If the nonlinearity of momentum space induces corrections to
the composition law of momenta, then the worldlines will have slightly different endpoints. So the interaction doesn't
appear local. Locality is recovered when the observer lays near $z^\mu$, that is, the interaction takes place near the origin of the coordinate system, so that $z^\mu \simeq 0$ and $x^\mu_j(\sigma_0) \simeq 0$ $\forall j$. This expresses the principle
of the relativity of locality.

To describe the physical picture perceived by different inertial observers, 
which are connected by (spacetime) translations and Lorentz transformations,
we need the Poisson brackets of dynamical
quantities with generators.
Then the transformation law of coordinates is
\begin{equation}
{x_j^\mu}' = x_j^\mu + a^\nu \left\{ {\mathcal K}_\nu(p^1,\dots p^n) , x_j^\mu \right\}~,
\end{equation}
and it is easy to prove that, at the level of the equations of motion,
this action effectively corresponds to translating \emph{classically} the
interaction point ${z^\mu}' = z^\mu + a^\mu$.

The translation generator in the case of more than one vertex
is not known, and neither is the Lorentz transformation generator,
even with a single vertex, if the momentum space is 
not simply a maximally symmetric space, where isometries are
homomorphisms of the composition law:
\begin{equation}
\Lambda (p \oplus q ) = \Lambda (p) \oplus \Lambda (q)~.
\end{equation}
From paper \cite{AFKS}~Êit is not clear
whether Lorentz transformations are a symmetry of the theory only
in this simple case or also in more general cases.

We are going to use the results of this Section on  the equations
of motion to find out how the Lorentz transformations look like in $\kappa$-Poincar\`e,
checking that they are indeed symmetries of the dynamics.

\section{$\kappa$-Poincar\'e representation on momentum space}
 
Hopf algebras possess in principle a sufficiently powerful structure to specify univocally
a manifold with a metric and a flat connection, which does not have to be necessarily 
the Levi-Civita one, because torsion and nonmetricity are allowed.

The bicrossproduct structure of $\kappa$-Poincar\'e, identified by Majid and Ruegg
allows one to distinguish between the translation sector, whose generators we call $P_{\mu}$, 
from the Lorentz sector, generated by boosts $N_{j}$ and rotations $R_{k}$.
%
The translation sector can be interpreted as the algebra of functions  over a manifold, which can be identified with the momentum space $\Sigma$, such that the generators $P_{\mu}$ assign coordinates to points on the manifold in a certain coordinate system according to
%
\begin{equation}
P_\mu (p) = p_\mu ~,
\end{equation}
where $p$ represents a point on the manifold, and $p_{\mu}$ its coordinates.

 A change of basis in the algebra generated by the $P_{\mu}$
 corresponds
to a change of coordinate system on the manifold:
\begin{equation}
P'_\mu = P'_\mu (P)~ \rightarrow ~~ P'_\mu(p') = p'_\mu  ~, ~~ \mathrm{where} ~~ p'_\mu = p'_\mu (p_{\nu}) ~.
\end{equation}
The coproduct  map is related to the composition
rule $\oplus$ of momentum space points,
\begin{equation}
\Delta P_\mu ( p , q ) = (p \oplus q)_\mu  ~.
\end{equation}
Then from the coassociativity axiom of Hopf algebras:
\begin{equation}
(\Delta \otimes \mathrm{id} ) \circ \Delta = (\mathrm{id} \otimes \Delta ) \circ \Delta ~,
\end{equation}
the associativity of the momentum composition rule follows,
\begin{equation}
((p \oplus q) \oplus k) = (p \oplus (q \oplus k)) ~,\label{eq:Associativity}
\end{equation}
which, in turn, implies the flatness of the connection on the momentum manifold, cf. Eq. \eq{eq:CurvatureDef}.
The counit  can be used to
identify the coordinates of 
the origin of momentum space $ \underline{0}$
\begin{equation}
P_\mu (\underline{0} ) = \varepsilon (P_\mu)~,
\end{equation}
in a way that is compatible with the antipode axiom   ($\mu$ is the multiplication of the Hopf algebra) 
\begin{equation}
\mu\circ (S \otimes \mathrm{id} ) \circ \Delta = \mu\circ ( \mathrm{id}  \otimes S) \circ \Delta  = \id ~ \varepsilon ~,
\end{equation}
if we relate the antipode $S$ with the inversion $\ominus$ in the following way:
\begin{equation}
S(P_\mu)( p ) =( \ominus p)_{\mu} ~.
\end{equation}
In fact in this way  we have
\begin{equation} 
( (S \otimes \mathrm{id} )  \circ \Delta )P_\mu (p,p) =( (\mathrm{id} \otimes S )  \circ \Delta )P_\mu (p,p)= ((\ominus p) \oplus p)_\mu =\varepsilon(P_\mu) ~.
\end{equation}
 
\begin{table}[htdp]
\begin{center}
\vspace{12pt}
\begin{tabular}{cc}
\hline
\hline
Hopf algebra $\mathcal H$ ~~ & ~~ Momentum space $\Sigma$
\\
\hline
$\Delta : \mathcal{H} \to \mathcal{H}  \times \mathcal{H} $ & $\oplus : \Sigma \times \Sigma \to \Sigma$ \\
$S : \mathcal{H}  \to \mathcal{H} $ & $\ominus : \Sigma \to \Sigma$ \\
$\varepsilon : \mathcal{H} \to \mathbb{R}$ &  $ \underline{0}: \mathbb{R} \to \Sigma $\\
generators $P_\mu$  & coordinate system $p_\mu$\\
change of basis & diffeomorphism \\
\hline
\hline
\end{tabular}
\caption{Duality between Hopf algebra and momentum space structures.\label{Tabellina}}
\end{center}
\end{table}
We end up having a neat picture relating the geometric structures on momentum space  introduced in Ref. \cite{AFKS}, and the algebraic structure of the $\kappa$-Poincar\'e Hopf
algebra, which we summarize in Table 1. 
The reason why we were able to get this is simple: the momentum composition rule $\oplus$, together with the origin $\underline{0}$ and the bilateral inverse $\ominus p$ equips the momentum space with an algebra loop structure: a group without the associativity axiom. If $\oplus$ is associative, then we have a group. And in particular we have a Lie group, because its elements are points on a manifold.
Now, it is well known \cite{Woronowicz} that abelian Hopf algebras are
dual structures to Lie groups, and they are introduced as algebras of functions over
the group. The duality allows to reconstruct everything about the group from the Hopf algebra and vice versa \footnote{With Hopf algebras, due to the coassociativity axiom, we are able
only to describe momentum spaces with flat connections. If we wanted to find the algebraic structure associated to a momentum space with a non-associative composition law, we would have had to rely on Hopf quasigroups \cite{HopfQuasigroups}.}.

\section{Geometric properties of the $\kappa$-Poincar\'e Hopf algebra}\label{GeomProperties}

In the previous Section we have shown the relation between the structures of the $\kappa$-Poincar\'e Hopf algebra and the ones of the associated momentum space.  So now we can deduce the physical properties of particles living on this momentum space according to the framework of Relative Locality outlined in Section 2.
To do this we need to describe in more details the geometric properties of the  momentum space associated to the $\kappa$-Poincar\'e algebra, specifying the metric, which allows to deduce the dispersion relation of particles (see Eq. \eq{eq:DispRelAFKS}), and the connection, with its nonmetricity and torsion, which are instead related to  particle interactions (see eqs. \eq{ConnectionAtAPoint}, \eq{eq:torsionAFKS} and \eq{defNonmetricity}).


For simplicity we will restrict the calculations  to the 1+1 dimensional version of the $\kappa$-Poincar\'e algebra.
The generalization to 3+1 dimensions is straightforward (and is discussed in Section \ref{sec:3+1}).
To fix notation we report the main properties of $\kappa$-Poincar\'e in the bicrossproduct basis. The generators satisfy the commutation rules
\begin{equation}
{[}P,E]= 0~, \qquad {[}N, P]=  \frac{\kappa}{2}(1 - e^{- 2 E /\kappa})  - \frac{1}{2\kappa}  P^2 ~, \qquad {[}N,E] = P ~, \label{1+1BicrossRelationsAlgebra}
\end{equation}
where $E$ and $P$ are the translation generators, and $N$ is the boost generator. The coalgebra is
\begin{equation}
\begin{array}{l}
\Delta E = E \ot \id + \id \ot E~, ~~~ \Delta P = P \ot \id + e^{-E/\kappa} \ot P~, \\
\Delta N  =  N \otimes \id + e^{-E/\kappa} \otimes N~,
\end{array} \label{1+1BicrossRelationsCoalgebra}
\end{equation}
and, finally, the antipodes and counits are
\begin{eqnarray}
 &&S(E) = - E ~, ~~~ S(P) = - e^{E/\kappa}  P ~,  ~~~ S(N) = -e^{E/\kappa}N ~, \\
 &&\varepsilon(E) = \varepsilon(P)  = \varepsilon(N) = 0 ~.
\end{eqnarray}

\subsection{Metric}

It was stated several times in the literature that $\kappa$-Poincar\'e describes a curved momentum manifold \cite{MajidRueggBicross,Majid-AlgebraicApproach}, and this manifold has been claimed to be a de Sitter space of radius $\kappa$ (see, in particular,  \cite{dSKow02,dSKow03,dSKow04}). Here we conclusively demonstrate that the metric
is  indeed that of a de Sitter space\footnote{During the final stages of preparation of this work we became
aware, through a talk given by L. Smolin at the
meeting Loops11, of an ongoing related project \cite{GACKowalski}
which reached similar conclusions about the metric
and connection for the $\kappa$-Poincar\'e Hopf algebra.}, but  in the next Sections we also show that the momentum
space is not simply de Sitter, because it is endowed with torsion and nonmetricity, that change
the connection in such a way that the curvature tensor is zero, unlike what happens
in a de Sitter space with the Levi-Civita connection.

To find the metric we  observe that the de Sitter line element in
comoving coordinates
\begin{equation}
ds^2 = d E^2 - e^{2E/\kappa} dp^2  \label{dScomoving}
\end{equation}
is invariant under the action of the $\kappa$-Poincar\'e
boosts \eq{1+1BicrossRelationsAlgebra}.
In fact we can exponentialize the action of the boost generators on the momenta,
in order to obtain the finite Lorentz transformations, as done in \cite{Rossano} (also see \cite{AmelinoCamelia:2000mn}):
\begin{eqnarray}
E' &=& E + \kappa ~ \log \left[ \left( \cosh \xi/2  + \frac{p}{\kappa} \sinh \xi/2 \right)^2 - e^{- 2 E / \kappa} \sinh^2 \xi/2  \right] \nonumber ~,
\\ \label{FiniteLorentzTransforms} \\
p' &=& \kappa \frac{\left( \mathrm{ch} \, \xi/2 + \frac{p}{\kappa} \mathrm{sh} \, \xi/2 \right) \left( \mathrm{sh} \, \xi/2 + \frac{p}{\kappa} \mathrm{ch} \, \xi/2 \right) - e^{-2 E /\kappa} \mathrm{ch} \, \xi/2 ~ \mathrm{sh} \, \xi/2 }{\left( \mathrm{ch} \, \xi/2 + \frac{p}{\kappa} \mathrm{sh} \, \xi/2 \right)^2 - e^{- 2 E / \kappa} \sinh^2 \, \xi/2 } 	\nonumber ~.
\end{eqnarray}
Then, plugging these expressions into the line element \eq{dScomoving} one
verifies that it is invariant:
\begin{equation}
ds'^2 = d E'^2 - e^{2E'/\kappa} dp'^2 ~.
\end{equation}

We can also show that the metric is de Sitter in a constructive way, which will also provide a useful coordinate system to do the computations in the following Subsection. 
Consider the change of basis (remember that a change of basis in the algebra corresponds to a change of coordinates on the momentum manifold):
\begin{equation}
\begin{array}{c}
\eta_0 = \kappa \sinh (E/\kappa) + e^{E/\kappa} P^2 /2\kappa ~,\vspace{6pt}\\
\eta_1 =  e^{E/\kappa} P ~.
\end{array}
\end{equation}
In this basis the algebra reduces to the Poincar\'e algebra
\begin{equation}
[\eta_0,\eta_1] = 0 ~, ~~~Ê [N,\eta_0] = \eta_1 ~, ~~~Ê[N,\eta_1] = \eta_0~,
\end{equation}
but the transformation is not 1 to 1, because it can be inverted in two ways
:
\begin{equation}
E_\pm = \kappa \log \left(\frac{\eta_0 \pm  \sqrt{\kappa^2 + \eta_0^2 - \eta_1^2}}{\kappa}\right)  ~, ~~~ÊP_\pm = \frac{\kappa ~ \eta_1}{\eta_0 \pm  \sqrt{\kappa^2 + \eta_0^2 - \eta_1^2}} ~. \label{TransformEtaP}
\end{equation}
This implies that the coalgebra in the new basis doesn't close,
\begin{equation}
\begin{array}{c}
\Delta \eta_1 = \eta_1 \otimes e^{E/\kappa} + \id \otimes \eta_1 	\vspace{6pt} \\
\Delta \eta_0 = \eta_0 \otimes e^{E/\kappa} + e^{-E/\kappa} \otimes \eta_0 + \frac{1}{\kappa}  e^{-E/\kappa} \eta_1 \otimes \eta_1 Ê~,\end{array}
\end{equation}
because we are not able to express  the $e^{E/\kappa}$ factor  in a unique way as a function of
$\eta_0$ and $\eta_1$.

However, if we introduce now the additional coordinate:
\begin{equation}
\eta_4 =   \kappa \cosh (E/\kappa) - e^{E/\kappa} P^2 /2\kappa~,
\end{equation}
we see that, since $\eta_4$ can be both positive and negative, it makes the 
change of basis invertible in a unique way:
\begin{equation}
E = \kappa \log \left(\frac{\eta_0 + \eta_4}{\kappa}\right)  ~, ~~~ÊP = \frac{\kappa ~ \eta_1}{\eta_0 + \eta_4} ~,
\end{equation}
and the coalgebra then closes:
\begin{equation}
\begin{array}{c}
\Delta \eta_0 = \frac{1}{\kappa}
\eta_0 \otimes (\eta_0 + \eta_4) + \frac{\kappa}{\eta_0 + \eta_4}
\otimes \eta_0 +  \frac{\eta_1}{\eta_0 + \eta_4}  \otimes \eta_1 Ê~, \vspace{6pt}\\
\Delta \eta_1 = \frac{1}{\kappa} \eta_1 \otimes (\eta_0 + \eta_4) + \id \otimes \eta_1 ~,	\vspace{6pt} \\
\Delta \eta_4 = \frac{1}{\kappa}
\eta_4 \otimes (\eta_0 + \eta_4) - \frac{\kappa}{\eta_0 + \eta_4}
\otimes \eta_0 -  \frac{\eta_1}{\eta_0 + \eta_4}  \otimes \eta_1 Ê~.
\end{array}
\end{equation}

\begin{figure}[!h]
\begin{center}
\includegraphics[width=0.6\textwidth]{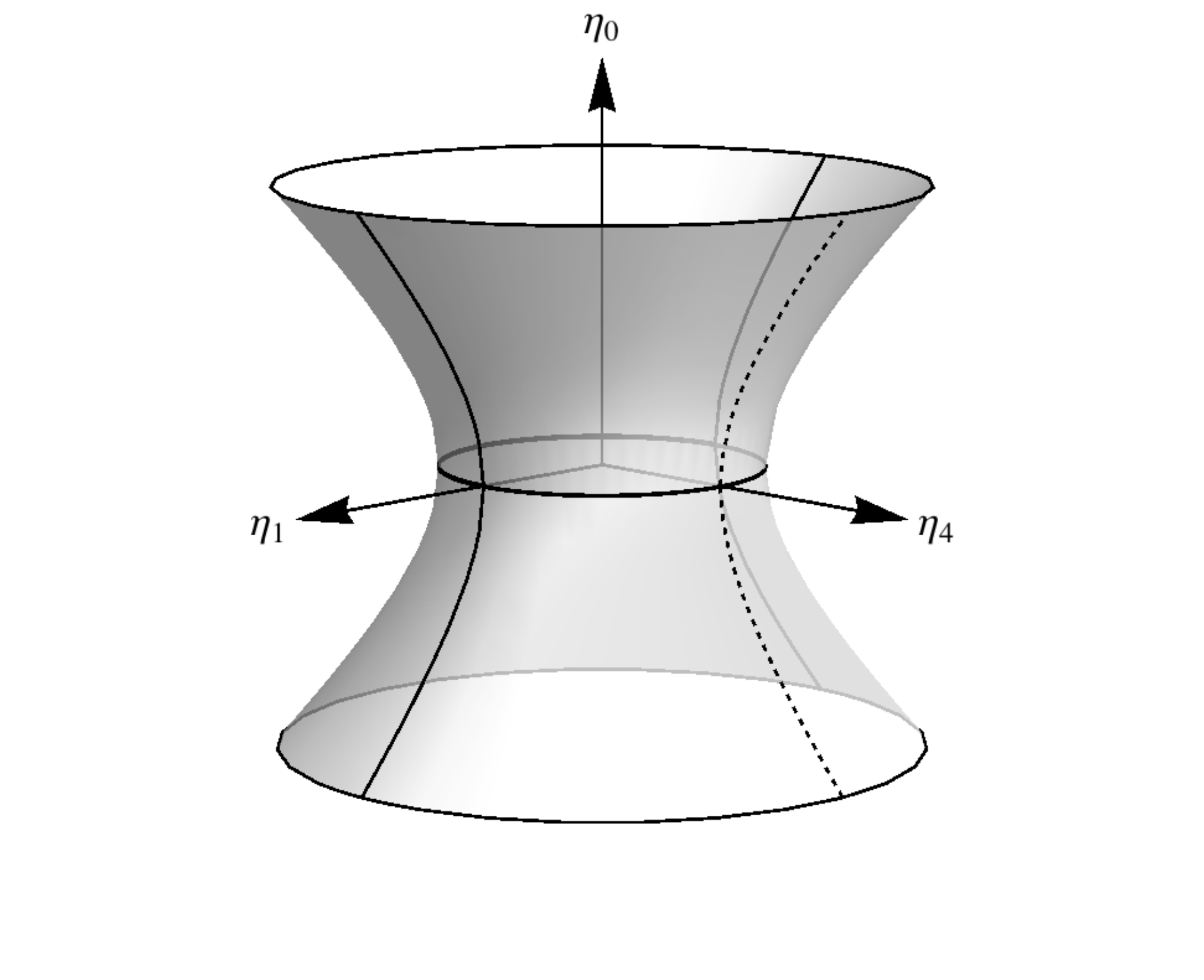}
\caption{de Sitter space, with the $\eta_a$ embedding coordinates.}
\end{center}\label{dSfig1}
\end{figure}

We are then able to recognize the  $\eta_a$ ($a=0,1,4$) generators\footnote{The coordinates $\{\eta_0,\eta_1,\eta_4\}$ were first introduced in \cite{dSKow02,dSKow03,dSKow04}, where a relation 
 between $\kappa$-Poincar\'e and de Sitter space was first conjectured.} as the ones associated (from a momentum space perspective) to the embedding
coordinates of a two-dimensional de Sitter space of radius $\kappa$, since they satisfy the constraint:
\begin{equation}
\eta_0^2 - \eta_1^2 - \eta_4^2 = - \kappa^2 ~. \label{Hyperboloid}
\end{equation}
It is also possible to show that then  $N$ is the generator of the Lorentz subalgebra $so(1,1) \in so(2,1)$ of the
isometries of the space:
\begin{equation}
 [N,\eta_0] = \eta_1 ~, ~~~Ê[N,\eta_1] = \eta_0~, ~~~Ê[N,\eta_4] = 0~.
\end{equation}

If now we induce the metric on the manifold defined by Eq. \eq{Hyperboloid} from
the flat metric in the ambient space, and we change back to the $\{E,p\}$ coordinates
according to \eq{TransformEtaP}, we recover the metric \eq{dScomoving}.


\subsection{Geodesics and particles dispersion relation}

\begin{figure}[!h]
\begin{center}
\includegraphics[width=0.6\textwidth]{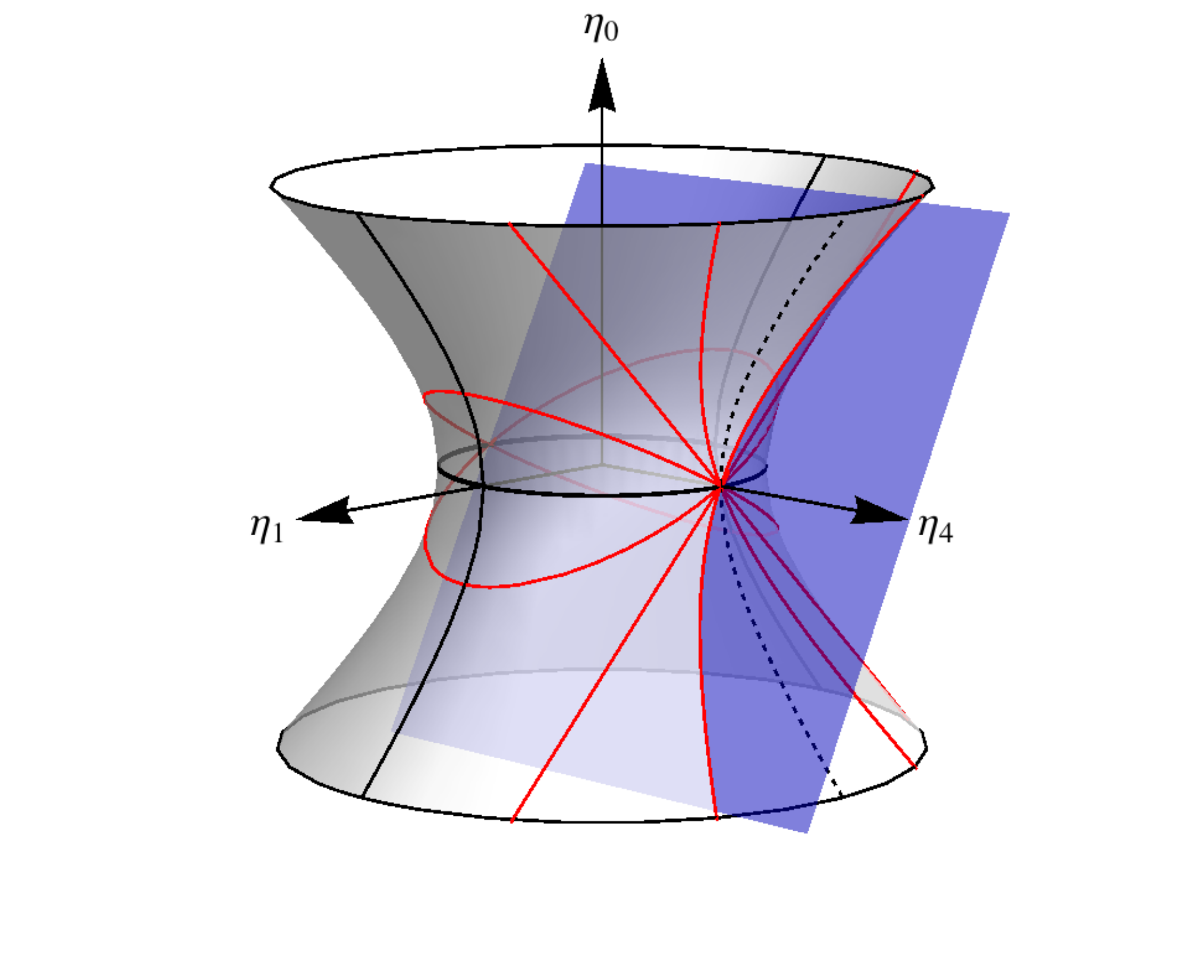}
\caption{Geodesics (in red) from the origin of the $\eta_a$ coordinate system in de Sitter space. They are given by the intersections with planes through the $\eta_4$ axis (in blue).\label{dSfig2}}
\end{center}
\end{figure}

Now that we have the metric of the momentum space associated to the $\kappa$-Poincar\'e algebra, we can derive the physical properties of particles living on this momentum space, studying its geodesic equation, the connection, the torsion and the nonmetricity.

In the Relative Locality framework, the mass of a particle is given by the geodesic distance of the particle's representing point on momentum space from the origin. So each particle with mass $m$ will live on a curve of constant geodesic distance from the origin, and the equation \eq{eq:DispRelAFKS} relating mass and geodesic distance gives the particle's dispersion relation.

The geodesics in a de Sitter space are easily obtained in the embedding coordinates. They are given \cite{deSitterExplained} by the intersection of the hyperboloid \eq{Hyperboloid} with the planes passing through the center (in embedding coordinates: $\{\eta_{0},\eta_{1},\eta_{4}\}\equiv\{0,0,0\}$).

To write the dispersion relation for particles living on this de Sitter momentum space we need  the geodesics that cross the origin $\underline 0$, which in the $\eta_a$ coordinates is the point $\{0,0,\kappa\}$ \footnote{Note how the essential role of the counit is here manifest: we know that the origin in the $\eta_a$ coordinate system has these coordinates because $\varepsilon(\{\eta_{0},\eta_{1},\eta_{4}\}) = \{0,0,\kappa\}$.}. So all the geodesics we are interested in are given by the intersections with the planes that contain the $\eta_4$ axis (see Fig. 2).

\begin{figure}[!h]
\begin{center}
\includegraphics[width=0.6\textwidth]{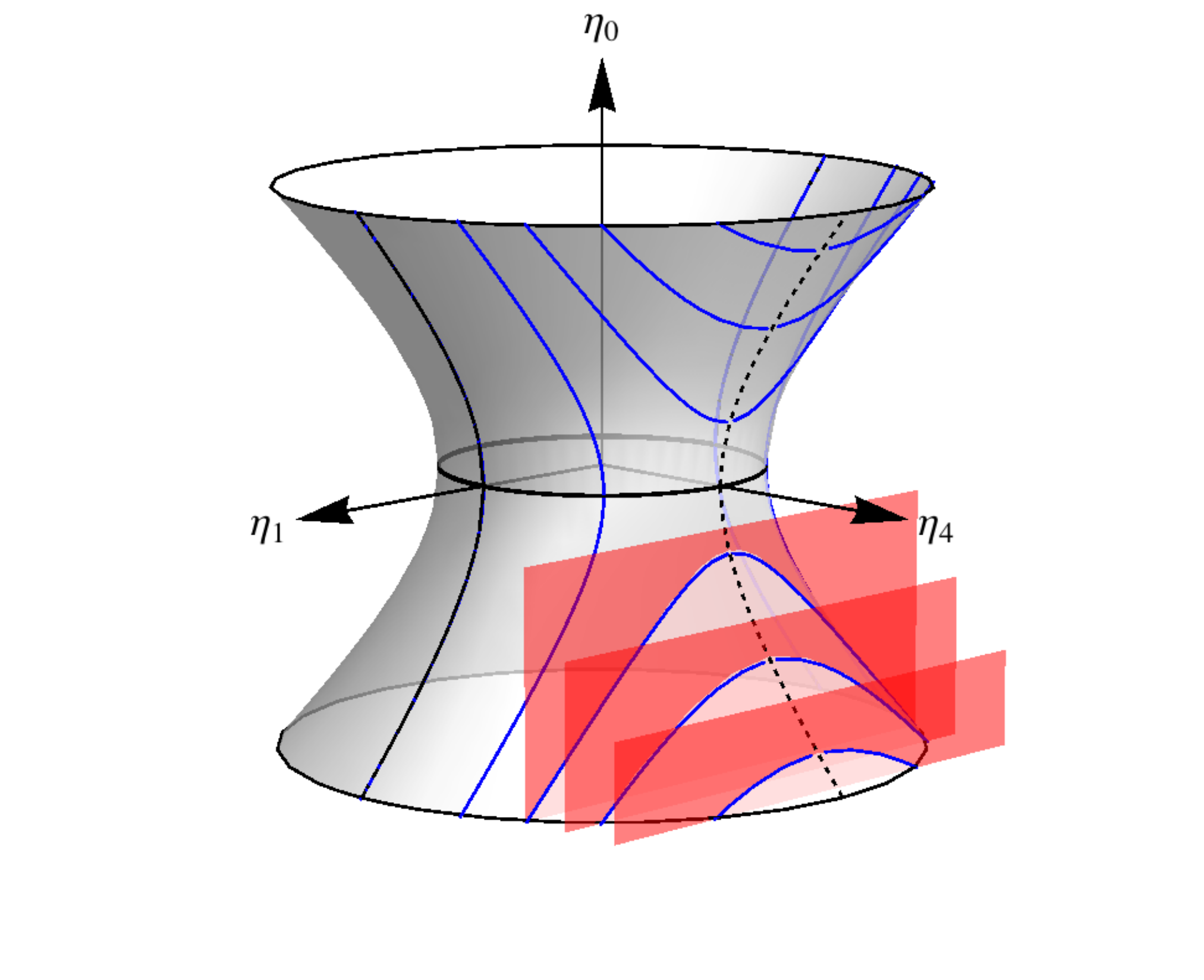}
\caption{Curves of constant geodesic distance (in blue) from the origin of the $\eta_a$ coordinate system. They are given by the intersections with the planes orthogonal to $\eta_4$ (in red).\label{dSfig3}}
\end{center}
\end{figure}

The curves with constant geodesic distance from the origin are obtained
 through the intersection with the planes that are orthogonal to the $\eta_4$ axis (See Fig. 3). Their equation in embedding coordinates is:
\begin{equation}
\eta_4 = \kappa \cosh ( d /\kappa )~,
\end{equation}
where $d$ is the (constant) geodesic distance of the curve. This can be seen by restricting
oneself to the plane $\eta_1 = 0$. The equation of the curve obtained by intersecting the 
de Sitter space with that plane is $\eta_4^2 - \eta_0^2 = \kappa^2$. This curve can be 
trivially parametrized by arc length as $(\eta_4 , \eta_0)= \kappa \, (\cosh \, s , \sinh \, s)$,
and the relationship between the dimensionless arc length and the geodesic distance
is $d = s \, \kappa$, in a de Sitter space of radius $\kappa$.

Then, in the bicrossproduct coordinates $E$, $P$, the equation satisfied by constant geodesic distance curves is:
\begin{equation}
d(E,P) = 
\kappa ~ \mathrm{arccosh}  \left(  \cosh (E/\kappa) - e^{E/\kappa} P^2 /2\kappa^2 \right)  ~.
\end{equation}
So, according to the Relative Locality construction, this should be taken as the dispersion relation of particles whose momentum space has coordinates and isometries described by the  $\kappa$-Poincar\'e algebra.

Notice that the usual proposal for the dispersion relation of $\kappa$-Poincar\'e is based on the Casimir \cite{LukierskiInventsKpoincare2,GiovanniMajidWaves,Nopure} :
\begin{equation}
\square_\kappa = 4 \kappa^2 \sinh^2 (E/2\kappa) - e^{E/\kappa} P^2 \equiv  2 \kappa  (\eta_4 -\kappa) ~, \label{UsualCasimir}
\end{equation}
which is  a nonlinear function of our geodesic distance. The difference can be reabsorbed into a nonlinear redefinition of the mass.

An interesting observation following from this analysis is that the geodesic distance naturally
selects a definition of mass as the rest energy of a particle. In fact, according to Eq. \eq{eq:DispRelAFKS}, the mass satisfies the relation:
\begin{equation}
\cosh (m/ \kappa ) =  \cosh (E/\kappa) - e^{E/\kappa} P^2 /2\kappa^2 ~,
\end{equation}
so that when $P=0$ the dispersion relation gives $E = m$, and when $\kappa \to \infty$
the relation reduces to $E^2-P^2 = m^2$.
If instead, as it was customary to do in literature until now \cite{LukierskiInventsKpoincare2,GiovanniMajidWaves,Nopure}, one uses the Casimir (\ref{UsualCasimir}) as the definition of the dispersion relation, then the rest energy and the mass would be related in a nonlinear way, $4 \kappa^2 \sinh^2 \left( \frac{E}{2 \kappa} \right)  = m^2$.


\subsection{Connection, torsion, nonmetricity and composition of particles momenta}

In Section 3 we have  derived  the properties of momenta composition rules from the properties of $\kappa$-Poincar\'e translation generators. On the other hand, in Section 2 we have stated that in the Relative Locality framework  momenta composition rules are related
to the geometric properties (connection, torsion) of the momentum space.
Here we show explicitly this relation for the $\kappa$-Poincar\'e momentum space.

From the co-associativity of the coproduct of the $\kappa$-Poincar\'e generators, which
means that the composition rule of momenta is associative (see Eq. \eq{eq:Associativity}),
it follows that the curvature vanishes\footnote{The
associativity of the composition law $\oplus$ trivially implies that of
the ``translated'' law $\oplus_k$, so the curvature vanish everywhere, according
to Eq. \eq{eq:CurvatureDef}.}.

The coproduct of the $P$ and $E$ generators, Eq. \eq{1+1BicrossRelationsCoalgebra}, can be used to write explicitly the ``translated'' composition law (\ref{TranslatedComposition})
\begin{equation}
\begin{array}{l}
(p \oplus _k q )_0=  p_0+ q_0- k_0 ~, \vspace{6pt}\\
(p \oplus _k q )_1= p_1 + e^{(k_0-p_0)/\kappa }(q_1- k_1)Ê~,
\end{array}
\end{equation}
which is needed to calculate  the connection at an arbitrary point as in Eq. \eq{ConnectionAtAPoint}.
Then the expressions of the connection and the torsion are
\begin{equation}
\Gamma^{\mu\nu}_\rho = - \left. \frac{\partial}{\partial p_\mu}\frac{\partial}{\partial q_\nu}(p \oplus _k q )_\rho \right|_{p=q=k}=  \frac{1}{\kappa}Ê~ {\delta^\mu}_0{\delta^\nu}_1 {\delta^1}_\rho~, \label{KappaConnection}
\end{equation}
\begin{equation}
T^{\mu\nu}_\rho (k) = \frac{1}{\kappa}Ê~ {\delta^{[\mu}}_0{\delta^{\nu]}}_1 {\delta^1}_\rho~.
\end{equation}
From the connection and the metric we can derive the  nonmetricity:
\begin{equation}
\begin{array}{l}
\nabla^\rho g^{\mu\nu} = \partial^\rho g^{\mu\nu} + \Gamma^{\mu \rho}_\sigma g^{\sigma\nu} + \Gamma^{\nu \rho}_\sigma g^{\mu \sigma}
\vspace{6pt}\\= -\frac{1}{\kappa } \bigg(2\, \delta ^{\mu }{}_1\delta ^{\nu }{}_1\delta ^{\rho }{}_0+\delta ^{\mu }{}_0\delta ^{\nu }{}_1\delta ^{\rho }{}_1+\delta ^{\mu }{}_1 \delta ^{\nu }{}_0\delta ^{\rho }{}_1\bigg)e^{2 E/\kappa }~.
\end{array}
\end{equation}
As already noticed in Section 3, the connection is flat, in the sense that the Riemann
tensor vanishes, due to the associativity of the composition law.


\section{Lorentz transformations}


%
The translation sector of $\kappa$-Poincar\'e can be interpreted as the algebra of functions over a curved momentum space,
while, as we have shown  in subsection 4.1,  the Lorentz sector generates a subalgebra of isometries on the momentum space.

We have also seen that we can  state a correspondence between the de Sitter momentum space defined by $\kappa$-Poincar\'e and the physical properties of particles living on it, but it is still not clear if the isometries on the momentum space represented by the boost generator actually correspond to transformations leaving the dynamics invariant. In particular the boost transformations need to be covariant also with respect to the composition of momenta.

In this Section  we will be actually able to find this covariant action of boosts on composed momenta.
A poorly known  ``back-reaction''  of the momenta on the Lorentz sector, found by Majid in \cite{Majid-AlgebraicApproach}, is the key to find this action. 

Let's define the boost transformations in momentum space (in the bicrossproduct basis) as in Eq. \eq{FiniteLorentzTransforms}
\begin{equation*}
\Lambda(\xi, p) =
\left( \begin{array}{c} p_0 + \kappa ~ \log \left[ \left( \cosh \xi/2  + \frac{p_1}{\kappa} \sinh \xi/2 \right)^2 - e^{- 2 p_0 / \kappa} \sinh^2 \xi/2  \right]
\\
\kappa \frac{\left( \mathrm{ch} \, \xi/2 + \frac{p_1}{\kappa} \mathrm{sh} \, \xi/2 \right) \left( \mathrm{sh} \, \xi/2 + \frac{p_1}{\kappa} \mathrm{ch} \, \xi/2 \right) - e^{-2 p_0/\kappa} \mathrm{ch} \, \xi/2 ~ \mathrm{sh} \, \xi/2 }{\left( \mathrm{ch} \, \xi/2 + \frac{p_1}{\kappa} \mathrm{sh} \, \xi/2 \right)^2 - e^{- 2 p_0 / \kappa} \sinh^2 \, \xi/2 }  \end{array} \right) ~,
\end{equation*}
where $\xi$ is the rapidity. Of course, since these transformations  preserve the metric (cf. subsection 4.1), they also leave invariant the geodesic distance 
\begin{equation}
d(\Lambda(\xi,p),\Lambda(\xi,q)) = d(p,q) ~.\label{eq:distanceinvariance}
\end{equation}
Moreover they  close an abelian group\footnote{This is the only point in which the 3+1 dimensional case shows some complications
with respect to the 1+1-d, because the Lorentz group in the 3+1-d case is nonabelian.
However, there are no novelties with respect to special relativity here because
the Lorentz subgroup is classical, and $\kappa$ won't intervene in the composition
law for rapidities},
\begin{equation}
\Lambda( \xi_1 , \Lambda(\xi_2 , p ) ) = \Lambda( \xi_1 + \xi_2 ,p) \label{eq:composition1+1} ~,
\end{equation}
and they reduce to ordinary Lorentz transformations in the limit $\kappa \to \infty$:
\begin{equation}
\Lambda(\xi, p) =
\left( \begin{array}{c} p_0 \cosh \xi + p_1 \sinh \xi
\\
 p_1 \cosh \xi + p_0 \sinh \xi
\end{array} \right) ~.
\end{equation}

We want to find how these transformations act on composed momenta: 
this  allows to determine how the momenta of various particles that interact in
a vertex would appear to a boosted observer. The trivial solution,
valid in special relativity, that each momentum transforms independently from the others,
of course doesn't work here, since
\begin{equation}
\Lambda( \xi,p\oplus q) \neq \Lambda(\xi,p) \oplus \Lambda(\xi,q) ~.
\end{equation}

A solution to this problem comes if we exploit this relation found  in \cite{Majid-AlgebraicApproach}:
momenta on which finite Lorentz transformations act turn out to have a ``back-reaction'' on them,
since they change the rapidity in a momentum-dependent way, that is compatible with the coproduct of momenta, and with the action
of Lorentz transformations on momenta themselves. This ``back-reaction'' is defined as the right action $\triangleleft : \mathbb{R} \times \Sigma \to \mathbb{R}$, that in bicrossproduct coordinates reads
\begin{equation}
\xi \triangleleft p = 2 \, \mathrm{arcsinh} \left( \frac{e^{-p_0/\kappa} \sinh \frac{\xi }{2} }{\sqrt{\left( \cosh \frac{\xi }{2}  + \frac{p_1}{\kappa} \sinh \frac{\xi }{2} \right)^2 - e^{-2 p_0/\kappa} \sinh^2 \frac{\xi }{2}}} \right) ~.
\end{equation}

This equation allows us to write the Lorentz transformation of 
 two composed  momenta as
\begin{equation}
\Lambda (\xi, q \oplus k ) =
\Lambda (\xi, q) \oplus \Lambda (\xi \triangleleft q, k ). \label{LorentzTrasfComposedMomenta}
\end{equation}
Then, if we call $q'$ and $k'$ the boosted momenta,
\begin{equation}
(q \oplus k )' = q' \oplus k'Ê~, ~~~ q' = \Lambda(\xi,q) ~, ~~k' =  \Lambda (\xi \triangleleft q, k ) ~,
\end{equation}
and this law ensures that both the transformed momenta, $q'$ and $k'$, are still on the mass-shell, because $k'$ is just boosted, even if with a $q$-dependent rapidity:
\begin{equation}
d(k',0) = d(\Lambda (\xi \triangleleft q, k ), 0 ) = d(k,0) ~.
\end{equation}
Eq. (\ref{LorentzTrasfComposedMomenta}) give a 
physical interpretation to the back-reaction, as a peculiar transformation law for the momenta
of particles interacting in a vertex. Each particle ends up transforming with a different rapidity,
and its rapidity depends on the momenta of the particle with which it interacts. 

Interestingly, the transformation law of any number of momenta participating to a
vertex is highly asymmetric with respect to the exchange of momenta, and it keeps
track of the order in which the momenta enter the vertex.

Considering the Lorentz transformation of  three composed momenta, and applying the Lorentz transformation in the two possible orders 
(thanks to the associativity of $\oplus$ we can forget about the brackets in the three-momenta sum)
\begin{eqnarray}
\Lambda (\xi, p \oplus q \oplus k ) &=&
\Lambda (\xi, p) \oplus   \Lambda ( \xi \triangleleft p, q) \oplus  \Lambda ( \xi \triangleleft p \triangleleft q, k)  \nonumber \\
&=&  \Lambda (\xi, p)  \oplus \Lambda (\xi \triangleleft p, q)  \oplus \Lambda (\xi \triangleleft p \oplus q , k  ) ~, 
\end{eqnarray}
we deduce that the associativity of $\oplus$ implies that the composition law of two consecutive actions of the momenta on the rapidity is:
\begin{equation}
\xi \triangleleft p \triangleleft q = \xi \triangleleft p \oplus q ~,
\end{equation}
expressing the covariance of the right-action of momenta on rapidities with 
respect to the momenta composition law.

A few remarks on the boosts $\Lambda(\xi,p)$ and the back-reaction $\xi \triangleleft p$
we have used. As observed in \cite{Majid-AlgebraicApproach} the boost and the
back-reaction are defined for every value of the rapidity only if the momentum lies within
the upper light cone $d(p,0) \geq 0$. Otherwise for every other $p$ there exists a finite critical boost $\xi_c$
that makes $p_0 \to - \infty$, and after which the transformation $\Lambda(\xi,p)$ is not
defined. Moreover, for every $\xi$ there exist a critical curve in momentum space, which lies outside of the
upper light-cone, on which $\xi \triangleleft p_c \to \pm \infty$, and after which the back-reaction is not defined.

In  \cite{Majid-AlgebraicApproach} a physical meaning is attributed to
this critical curve.  In fact, the commutation relations of the $\kappa$-Poincar\'e quantum group dual to the algebra are such that the commutator between translations and Lorentz transformations
has a singularity on the critical curve. That is interpreted as an infinite uncertainty
for certain states of this algebra. The physical meaning of generalizing to quantum operators the parameters of Lorentz transformations or translations connecting 
different inertial observers has not yet been clarified, therefore the meaning
of these infinite uncertainty states remains mysterious.

\begin{figure}[!h]
\begin{center}\label{dSfig4}
\includegraphics[width=0.6\textwidth]{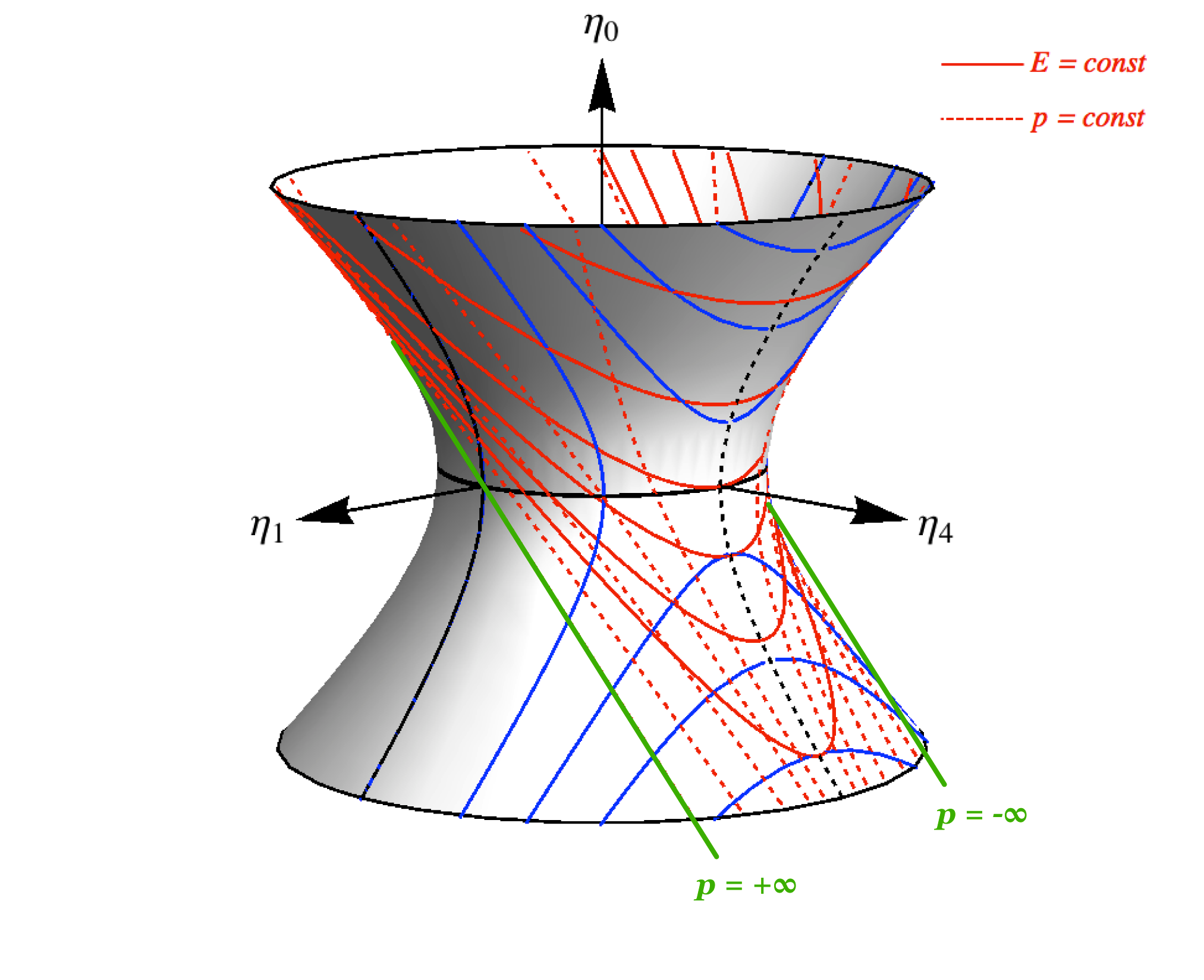}
\caption{The comoving coordinate system of de Sitter space. In solid red there are the $E =  const$ curves, and in dotted red the $p = const$ curves. The dispersion curves are in blue. In green the two critical curves corresponding to the $p = \pm \infty$ coordinate
singularity are shown. }
\end{center}
\end{figure}

The geometric interpretation of the bicrossproduct momentum space
suggests that the singularity encountered in \cite{Majid-AlgebraicApproach}  might be unphysical.
Here we want to remark that, in the geometric setting provided by Relative Locality,
the critical curve appears to be due only to a coordinate singularity, which is a well-known
property of the comoving coordinate system. In fact these coordinates only cover
half of the de Sitter space, and the $p$ coordinate diverges over the critical 
curves, shown in green in Fig. \ref{dSfig4}. The de Sitter space is cut in half by
the two critical curves, and another complementary set of coordinates is needed
to cover the other half. This feature of the bicrossproduct basis  has already been
noticed in \cite{dSKow02}.

So there appears to be nothing special with comoving coordinates,
being just a (possibly convenient) choice of coordinate system for a manifold.
It would be interesting to compute the dual to the $\kappa$-Poincar\'e Hopf
algebra in the ``embedding'' basis $\eta_a$, but we leave this for further studies.

An issue with the critical curves is, however, present: the two halves of de Sitter space
that are delimited by the two curves are closed under coproduct. This means that
a momentum lying in one of the two halves might have been generated only by
the combination of two momenta in the same half. However the two halves are not
closed under Lorentz transformations, and one can move any momentum from one half
to the other. This has led to speculations regarding a possible breakdown of Lorentz
invariance in $\kappa$-Poincar\'e \cite{FreidelJurekLorentzBreaking}.
We refer the reader to the most recent discussion of the issue, and its possible solution \cite{MicheleJurek}.

\section{Equivalence between inertial observers}

In the previous Section we have seen that when applied to interacting particles, Lorentz transformations act differently on each particle. The rapidity with which they act on each single particle depends on
the momenta of the other particles which participate to the vertex and on their order.
Here we show that physics is left invariant by these kind of Lorentz transformations, showing that the equations of motion \eq{BulkEquations} and \eq{BoundaryEquations} for the particles in the vertex are invariant.

Let's consider a vertex with $n$ interacting particles whose momenta composition law is $\mathcal{K} =k^1 \oplus \dots \oplus k^n$, which is boosted with rapidity parameter $\xi$, and let's call
\begin{equation}
 \xi^j \equiv \xi \triangleleft k^1 \oplus \dots \oplus k^{j-1} ~,
\end{equation}
the rapidity with which the $j$-th moment boosts, so that
\begin{equation}
({k}^j)' =  \Lambda(\xi^j,k^j) ~,
\end{equation}
and the conservation law transforms as
\begin{equation}
\mathcal{K'} = \Lambda(\xi^1,k^1) \oplus \dots \oplus \Lambda(\xi^n,k^n) \equiv \Lambda(\xi, \mathcal{K}) ~.
\end{equation}
which means that it's invariant ($ \mathcal{K} = \underline{0} ~\RightarrowÊ\Lambda(\xi, \mathcal{K}) = \underline{0}$).
From this and from the invariance of the geodesic distance under boosts we see
that both the constraints (\ref{constraint1}) and (\ref{constraint2}) are invariant.

The particle coordinates will transform according to the transformation rule
of covectors under diffeomorphisms:
\begin{equation}
 \qquad ({x_{j}'})^\mu = {x}^\nu_j  \frac{\partial k^j_\nu}{\partial ({k}^j)'_\mu}  = {x}^\nu_j  \frac{\partial \Lambda( - \xi^i , (k^j)')_\nu}{\partial ({k}^j)'_\mu} ~.
\end{equation}
The bulk equations of motion \eq{BulkEquations}  are invariant under these transformations
\begin{eqnarray}
{\dot k}^j_\mu = 0  \Rightarrow  { ({\dot k}}^j)'_\mu =0~, \qquad \mathcal{K}_\mu = 0 \Rightarrow  \mathcal{K'}_\mu= 0~,  \nonumber
\\
\\
\qquad {\dot x_j}^\mu = N_j \frac{\partial D(k,0)}{\partial k^j_\mu}  \Rightarrow {(\dot{x}_j)'}^\mu = N_j \frac{\partial D(k',0)}{\partial ({k}^j)'_\mu} ~. \nonumber
\end{eqnarray}
The invariance of the first equation  is trivial: if all the $k^j$'s are constants, then also the transformed ones are so. The invariance of the second equation comes from the invariance of the origin, $\Lambda(\xi,\underline{0}) =\underline{0}$; and the last equation can be
verified by direct computation.

Also the boundary equations \eq{BoundaryEquations} are invariant
\begin{equation}
x_j^\mu(s_0) = z^\nu \frac{\partial \mathcal K_\nu}{\partial k^j_\mu} ~~Ê\Rightarrow ~~ {x'}_j^\mu(s_0) = {z'}^\nu \frac{\partial \mathcal K'_\nu}{\partial (k^j)'_\mu} ~,
\end{equation}
if the $z$'s transform as:
$$
{z'}^\mu = z^\nu  \frac{\partial {\mathcal K}_\nu}{\partial {\mathcal K'}_\mu} ~.
$$
Notice that  $\frac{\partial {\mathcal K}_\nu}{\partial {\mathcal K'}_\mu} = {\Lambda^\nu}_\mu$, and ${\Lambda^\nu}_\mu$ is the classical Lorentz transformation of rapidity $-\xi$\footnote{$\frac{\partial {\mathcal K}_\nu}{\partial {\mathcal K'}_\mu} $ is equal to $ \left. \frac{\partial p_\nu}{\partial \Lambda(\xi, p)_\mu} \right|_{p=0} $, which can be easily shown to be equal to a classical Lorentz transform.}.

So, interestingly, it turns out that the vertex coordinates $z^\mu$ transform classically both 
under translations and under Lorentz transformations.
Mathematically, this is a consequence of the fact  that both these transformations are
identical to the classical ones near the origin of momentum space\footnote{A fact that
we relate to the ``dual equivalence principle'' formulated in \cite{AFKS}, which states
that locally the geometry of momentum space is that of Minkowski.}, and the fact that
the $z^\mu$s transform under diffeomorphisms $p_\mu = f_\mu(p)$ as (see Ref. \cite{LeeLaurentGRB})
\begin{equation}
{z'}^\mu = z^\nu \left.  {\left[ \left( \frac{\partial f}{\partial p} \right)^{-1}\right]^\mu}_\nu \right|_{p=\underline{0}} ~,
\end{equation}
where $ \left.  {\left[ \left( \frac{\partial f}{\partial p} \right)^{-1}\right]^\mu}_\nu \right|_{p=\underline{0}} $ is the inverse
diffeomorphism \emph{calculated at the origin}.

A comment on Lorentz transformations between inertial observers.
In special relativity the rapidity of the boost is related to the velocity of one reference frame with respect to the other,
irrespective of the particle content of the system under consideration. In our framework the Lorentz transformations
are informed about that content. This doesn't prevent us to associate a rapidity to the observers: since the 
back-reaction of the momenta on the rapidity has the group property we have just shown, one can always
express the transformation in terms of the rapidity with which one chosen particle transform. An inertial system
has to be defined in an operational way, \emph{e.g.} Alice is in the reference system where particle $1$
is measured to have momentum $p_1$, and Bob is defined by having measured the momentum of particle $1$
to be $p_1' = \Lambda(\xi_1 , p_1)$. Then one can predict the value of the momenta $p'_2,\dots,p'_N$ of all the other 
particles in the process as measured by Bob, knowing their momenta  $p_2,\dots,p_N$ in Alice's frame.
Nothing will be left undetermined. An issue arises, however, if some of the particles in the process cannot
be measured. In that case one is not able to predict some of the $p'_2,\dots,p'_N$, an issue that is not present
in special relativity, where all of the momenta transform independently and with the same rapidity.
This is an additional difficulty which makes things significantly harder for the experimenter. 
Interestingly, this might turn out even to be an advantage over
ordinary special relativity: in special relativity one has only limited knowledge of the momenta of the particles that cannot be measured directly,
which is the knowledge coming from the kinematical constraints (i.e. missing energy and momentum). In the framework discussed in the paper,
one can deduce something more about the unobserved particles by the way the momenta of the observed particles transform under boost.

\section{$z_\mu$ coordinates and $\kappa$-Minkowski spacetime}

One can use the $\kappa$-Poincar\'e connection (\ref{KappaConnection})
to calculate the parallel transport along a geodesic of an infinitesimal vector $dq$,
living in the tangent space to the momentum space at the point $p$, from the
point $p$ to the origin \cite{AFKS},
\begin{equation}
(p \oplus dq)_\mu = p_\mu + dq_\nu {(\tau_L)^\nu}_\mu (p) ~,
\qquad  \tau_L (p) = \left( \begin{array}{cc} 1 & 0 \\ 0 & e^{- p_0 /\kappa} \end{array} \right)~, \label{ParallelTransport}
\end{equation}
where $\tau_L$ is the parallel transport matrix, which relates the components of $dq$
at $p$ to its parallely transported components at the origin.

In Ref. \cite{AFKS} the authors obtain the coordinates $z^\mu_j$ from the coordinates of the $j$-th particle with momentum $p^j$
(therefore living in the tangent space to the momentum space at the point $p$), by parallelly transporting
them along a geodesic toward the origin of the momentum space:
\begin{equation}
z^\mu_j \equiv x^\nu_j {(\tau_L)^\mu}_\nu (p^j)  ~.
\end{equation}
so that $z^0 = x^0 $ and $z^1 = x^1 e^{- p_0 /\kappa}$.
These coordinates are not canonical as the  $x^\mu_j$'s, that close canonical
Poisson brackets with the $p^j_\mu$:
\begin{equation}
 \{ x^\mu_j, p_\nu^k \} = {\delta^\mu}_\nu \, {\delta^k}_j ~.  
\end{equation}
They instead close a Lie algebra among them
\begin{equation}
\{ z^1_j, z^0_k \} = \frac{1}{\kappa} z^1_j  \, \delta_{jk} ~. \label{ZetaKappaMinkowski}
\end{equation}
This algebra is the same satisfied by the coordinates of $\kappa$-Minkowski space,
that  is expected to be the noncommutative spacetime
whose symmetries are described by $\kappa$-Poincar\'e (see  \cite{MajidRueggBicross,Sitarz,Nopure}).

Eq. (\ref{ZetaKappaMinkowski}) indicates that relative
locality may describe the ``$\hbar \to 0$ relics'' of  the 
$\kappa$-Minkowski noncommutative spacetime, which should be recovered
upon quantization,  transforming the Poisson brackets into commutators
and the coordinates $z^\mu_j$ into operators $\hat z^\mu_j$:
\begin{equation}
\{ z^1_j, z^0_k \} = \frac{1}{\kappa} z^1_j  \, \delta_{jk} ~~ \rightarrow ~~
[ \hat z^1_j, \hat z^0_k ] = i \, \lambda \, \hat z^1_j  \, \delta_{jk} \label{QuantumKappaMinkowski}
\end{equation}
where $\lambda = \hbar / \kappa$ is a length scale. This provides  a hint for the physical interpretation of the $\kappa$-Minkowski algebra, as an algebra of functions over a noncommutative spacetime.
One comment on the meaning of this ``quantization'': since in the Relative Locality framework
the creation and annihilation of particles are allowed, we expect that a
quantization of the model should involve second quantization methods,
with the introduction of a Fock space to represent multi-particle states.
The operators $\hat z^\mu_j$ in (\ref{QuantumKappaMinkowski}) still make
sense, in this setting, as the particle coordinates of a system of free particles,
in which the effects of particle creation and annihilation can be ignored.

\section{3+1 dimensional $\kappa$-Poincar\'e}
\label{sec:3+1}
Throughout the paper we have discussed the $1+1$ dimensional $\kappa$-Poincar\'e algebra. Here we show how to generalize to the more physical $3+1$-D case.

Let us start by noting down the $3+1$ dimensional $\kappa$-Poincar\'e algebra ($\mu,\nu=0,...,3$, $j,k,l=1,...,3$):
\begin{eqnarray}
&[P_{\mu},P_{\nu}]&=0,\quad [N_{j},P_{0}]=P_{j} ~,\nonumber\\
&[N_{j},P_{k}]&=\delta_{jk}\left(\frac{\kappa}{2}\left(1-e^{-2P_{0}/\kappa}\right)+\frac{1}{2\kappa}|\vec P|^{2}\right)-\frac{1}{\kappa}P_{j}P_{k}~, \nonumber \\
&[N_{j},N_{k}]&=-\epsilon_{jkl}R_{l},\quad [R_{j},P_{0}]=0,\quad [R_{j},P_{k}]=\epsilon_{jkl} P_{l}~,\nonumber\\
&[R_{j},N_{k}]&=\epsilon_{jkl}N_{l},\quad [R_{j},R_{k}]=\epsilon_{jkl}R_{l}~,
\end{eqnarray}
and coalgebra:
\begin{eqnarray}\label{3+1kappaPoincCoalg}
&\Delta P_{j}&=P_{j}\ot \id+e^{-P_{0}/\kappa}\ot P_{j},\quad\Delta P_{0}=P_{0}\ot\id+\id\ot P_{0}~, \nonumber\\
&\Delta N_{k}&=N_{k}\ot\id+e^{-P_{0}/\kappa}\ot N_{k}+\frac{1}{\kappa}\epsilon_{kjl}P_{j}\ot R_{l}~,\nonumber\\
&\Delta R_{j}&=R_{j}\ot\id+\id\ot R_{j}~,
\end{eqnarray}
and, finally, antipodes and counits
\begin{equation}
\begin{array}{l}
\varepsilon(P_\mu) = 0 ~,  ~~~ \varepsilon(N_j) = 0 ~, ~~~ \varepsilon(R_k) = 0Ê~, \vspace{6pt}\\
  S(P_0) = - P_0 ~, \qquad S(P_j) = - e^{P_0/\kappa}  P_j ~, \vspace{6pt}\\
S(N_j) = -e^{P_0/\kappa}N_j +\frac{i}{\kappa} \epsilon_{jkl} e^{\lambda P_0} P_k R_l ~, ~~~ S(R_k) = - R_k ~.
\end{array} \label{3+1kappaPoincAntipodeCounit}
\end{equation}
It is easy to check that the metric:
\begin{equation}
ds^{2}=dE^{2}-e^{2 E/\kappa}|d\vec p|^{2}
\end{equation}
is invariant under the infinitesimal Lorentz transformations (first order in the transformation parameters $\vec\xi$ and $\vec \theta$ associated, respectively, to the generators $\vec N$ and $\vec R$)
\begin{eqnarray}
E' &=& E +\vec p\cdot\vec \xi \nonumber ~, \\ ~\\
\vec p\,' &=& \vec p+\vec \xi\left( \frac{\kappa}{2}\left(1-e^{-2E/\kappa}\right)+\frac{1}{2\kappa}|\vec p|^{2}\right)-\frac{\vec p}{\kappa}(\vec\xi\cdot\vec p)	+\vec \theta\times \vec p\nonumber ~.
\end{eqnarray}

The embedding coordinates in the 3+1-dimensional case are a trivial generalization of Eq. ((30)) and ((34)),
\begin{equation}
\begin{array}{l}
\eta_0 = \kappa \, \sinh (P_0/\kappa) + e^{P_0/\kappa} \, |\vec P|^2/2\kappa  ~,  \vspace{6pt}\\
\vec \eta =  e^{P_0/\kappa} \, \vec P ~,  \vspace{6pt}\\
\eta_4 = \kappa \, \cosh (P_0/\kappa) - e^{P_0/\kappa} \, |\vec P|^2/2\kappa  ~,
\end{array}
\end{equation}
Because of the undeformed action of rotations, also the dispersion relation, the expression for the connection,
\begin{equation}
\Gamma^{\mu\nu}_\rho = \frac 1 \kappa \sum_{j=1}^3 {\delta^\mu}_0 \, {\delta^\nu}_j \, {\delta^j}_\rho ~,
\end{equation}
that of the torsion and  nonmetricity are the trivial generalization of the ones reported in section \ref{GeomProperties}, when one assumes invariance under rotations.

What requires more attention when generalizing to $3+1$ dimension is the composition of Lorentz transformations and the transformation of a system of particles. The composition of two transformations is modified because in $3+1$ dimensions the Lorentz group is nonabelian. 
The composition of two boosts n $3+1$ dimensions is different from a trivial generalization of equation (\ref{eq:composition1+1}) but the modification is the same as the one needed in special relativity when going from $1+1$ to $3+1$ dimensions (\emph{i.e.} there is no $\kappa$-dependent effect):
\begin{eqnarray}
\Lambda (\xi_{1},\Lambda(\xi_{2},p))=\Lambda(\xi_{1}\circ   \xi_{2},p)~, \\
 \xi_{1}\circ \xi_{2}=\frac{1}{1+\vec\xi_{1}\cdot\vec \xi_{2}}\left(\vec\xi_{1}+\vec\xi_{2}+\frac{\gamma_{\xi_{1}}}{1+\gamma_{\xi_{1}}}\vec\xi_{1}\times(\vec\xi_{1}\times\vec\xi_{2})\right)\,, \nonumber
\end{eqnarray}
where $\gamma_{\xi_{1}}=\frac{1}{\sqrt{1-|\vec\xi_{1}|^{2}}}$. A similar formula for two Lorentz transformations in
full generality (\emph{i.e.} including rotations) is not, to our knowledge, available.

Concerning the transformation of a system of particles under the Lorentz sector of $\kappa$-Poincar\'e algebra, we have seen in section 5 that we need to introduce a `back-reaction' of the momenta on the transformation parameters. The reason for this has to be traced back to the nontrivial form of the coproduct of the boost generator. In the following we discuss what happens in $3+1$ dimensions, limiting the discussion to the first order in the transformation parameters $\vec \xi$ and $\vec \theta$.

In $3+1$ dimensions, the coproduct of boost generators $N_{j}$ contains a rotation generator (see the coalgebra reported at the beginning of this section). So we do expect that  the back-reaction of momenta on the $\vec \xi$ transformation parameter generates also a rotation transformation. And indeed it is possible to show that the Lorentz transformation of a couple of particles with momenta $p$ and $q$ is:
\begin{equation}
\Lambda(\{\vec \xi,\vec\theta\},p\oplus q)=\Lambda(\{\vec \xi,\vec \theta\},p)\oplus \Lambda(\{\vec \xi,\vec\theta\} \triangleleft p, q),
\end{equation}¥ 
where
\begin{equation}\label{3+1backreactionExpression}
\{\vec \xi,\vec\theta\} \triangleleft p=\{e^{-p_{0}/\kappa}\vec \xi,\vec\theta-\frac{\vec\xi\times\vec p}{\kappa}\} \,.
\end{equation}
A few comments are in order.
\begin{enumerate}
\item Eq. (\ref{3+1backreactionExpression}) involves a back-reaction of the momenta on the \emph{rotations}. This
could come unexpected to some readers, as the common wisdom regarding $\kappa$-Poincar\'e is that the rotations
are undeformed. But Eq. (\ref{3+1backreactionExpression}) shows that, \emph{in presence of a boost}, the rotation 
parameter receives some back-reaction by the other momenta entering a vertex. Interestingly, even if the Lorentz
transformation acting on the left momentum contains no rotation part ($\vec \theta = 0$), then the back-reaction
induces a rotation of the right momentum with infinitesimal rotation vector $\frac 1 \kappa \, \vec \xi \times \vec p$.
This is the physical manifestation of the ``no-pure-boost'' principle discovered in \cite{Nopure}, and clarifies the way it is physically realized in processes.
\item  Notice how the first-order expression of the back-reaction (\ref{3+1backreactionExpression}) is closely related to
the coproduct of the boost and rotation generators (\ref{3+1kappaPoincCoalg}). The same happens in the 1+1 case,
where the back-reaction takes the form $ \xi \triangleleft p =   e^{-p_0/\kappa} \, \xi $ and the second term in the
boost coproduct is $ e^{-E/\kappa}  \otimes N$. The back-reaction is clearly a consequence of the coproduct
of the Lorentz generators. An explicit dictionary translating the Hopf-algebra structures of the Lorentz sector
into momentum-space structures like the back-reaction, on the lines of our Table~\ref{Tabellina} above, is clearly desirable. 
We won't comment further on the issue, leaving it for future work.
\end{enumerate}

\section{Conclusions and outlook}

The $\kappa$-Poincar\'e Hopf algebra has been subject to an intense study since
its discovery, almost 20 years ago. It attracted such a large interest because
it provides a treatable example of ``quantum geometry'',
described through the language of symmetries.
But until now a coherent picture had not been found, where its physical implications could
be unambiguously determined, and a connection with the experiments be made. 

In this paper we established Relative Locality as the natural paradigm in which
one should interpret the implications of $\kappa$-Poincar\'e for physics.
This paradigm comes with a simple and coherent physical model, which
allows to unambiguously determine the new effects that $\kappa$-Poincar\'e
implies, thus finally allowing for the long-sought connection with the experiments.


We determined that $\kappa$-Poincar\'e determines a Relative Locality model with
a momentum space which
metrically is de Sitter, with radius of curvature $\kappa$. This momentum space
has a non-metric connection with zero curvature, and non-zero torsion and nonmetricity.
%
%

These results lead to a fully workable model of point particles interacting with point-like,
but ``relatively-local'', interactions, with a non-symmetric and associative deformed conservation law of momentum.

We showed also the compatibility of the model with the relativity principle, and we found the form of
the Lorentz transformations for every system of particles, while previously this was known only for 
free particles. This allows us to confront, in real-world scenarios, the observations of different inertial
observers, a task  that was previously made impossible by the lack of a consistent law of transformation for
systems of interacting particles.
Interestingly, under boost transformations, on-shell momenta remain on-shell, with the same masses,
but when we apply boosts to several interacting particles, the rapidity with which  the momentum of each particle is boosted depends on the  momenta of the other particles taking part to the interaction.

To achieve our results, we showed the equivalence between the Hopf algebra
structures of $\kappa$-Poincar\'e and the geometric construction that realizes the
principle of Relative Locality. This construction is directly applicable also to
other Hopf algebras, and some analyses on the same lines are under development.

An observation: $\kappa$-Poincar\'e was initially obtained as the In\"on\"u-Wigner
contraction of another Hopf algebra, known as q-de Sitter \cite{Lukierski:1991pn,Lukierski:1991ff,Lukierski:1992dt,Ballesteros:2004eu}. This algebra depends
on two constants,   $H$ (dimensionful) and $q$ (dimensionless), since it has been
introduced as the $q$-deformation of the de Sitter algebra with radius $H^{-1}$. The contraction to $\kappa$-Poincar\'e corresponds to the limit $H\to 0$, $q \to 1$, where $H/ \log q \to \kappa$. One can trade $q$ for $\kappa = H/ \log q$, which plays the role of an ultraviolet constant, while $H$ is infrared \cite{Marciano:2010gq}. This suggests an interpretation
of q-de Sitter as the symmetry algebra of a system which possesses both curvature
in momentum space and in spacetime. An interesting question is whether this situation
can be fitted into the Relative Locality scheme.



\section*{Acknowledgments}

FM thanks L. Freidel for useful discussions during the early stages of this work. 

\section*{References}
\bibliographystyle{unsrt}
\bibliography{BibRelLocRevised}

\end{document}